\documentstyle[epsfig]{aipproc}

\begin{document}
\title{Ferroelectric effects in PZT}

\author{L. Bellaiche, J. Padilla and David Vanderbilt}
\address{Department of Physics and Astronomy,
Rutgers University, Piscataway, New Jersey 08855-0849}

%\lefthead{LEFT head}
%\righthead{RIGHT head}
\maketitle

\begin{abstract}
First-principles calculations are performed to investigate alloying and 
ferroelectric effects in lead zirconate titanate (PZT) with high Ti composition.
We find that the main effect of alloying in the paraelectric
phase of PZT is the existence of two sets of B--O bonds, i.e., shorter
Ti--O bonds {\it vs.} longer Zr--O bonds. 
On the other hand, ferroelectricity leads to the formation of
very short covalent Ti--O bonds and to the formation of covalent chains of
Pb--O bonds. The covalency in the ferroelectric phase is mainly 
induced by an enhancement of hybridization between Ti $3d$ and O $2p$, and
between Pb $6s$ and O $2p$. These hybridizations induce a striking decrease 
of the effective charges when going from the paraelectric to the
ferroelectric phase of PZT.
\end{abstract}

\section*{Introduction}

The technologically important lead zirconate titanate alloy (i.e.,
PbZr$_{1-x}$Ti$_{x}$O$_{3}$ usually denoted as PZT) has an 
interesting phase diagram \cite{Lines}. Increasing 
the Ti $x$ composition yields progressively
the following {\it ground state} phases: an antiferroelectric orthorhombic
phase for $x$ $\lesssim$ 0.1, a ferroelectric rhombohedral FE$_{2}$ phase for 
0.1 $\lesssim$ $x$ $\lesssim$ 0.4, another ferroelectric rhombohedral 
FE$_{1}$ phase for 
0.4 $\lesssim$ $x$ $\lesssim$ 0.5, and finally a tetragonal 
ferroelectric phase for $x$ larger 
than 50\%.
The high-temperature phase of this alloy at all compositions is the 
cubic perovskite structure. 

Previous theoretical studies \cite{Szabo} focused on the 
long-range B-site ordering effects 
in PZT for a Ti composition equal to 0.5, i.e., close to the morphotropic
phase boundary between rhombohedral and tetragonal ferroelectric phases. 
The subject of the present theoretical study is rather different:
we will investigate alloying and
ferroelectric effects on structural, chemical and dielectric
properties in the tetragonal phase of the PZT alloy.
In other words, we would like to know what are the effects of alloying and
ferroelectricity on bond lengths, chemical bonding and Born 
effective charges  in PZT with high Ti content.

\section*{Method}

\begin{figure} [b!] % fig 1
\begin{center} \leavevmode
\epsfxsize 1.48truein
\epsfbox{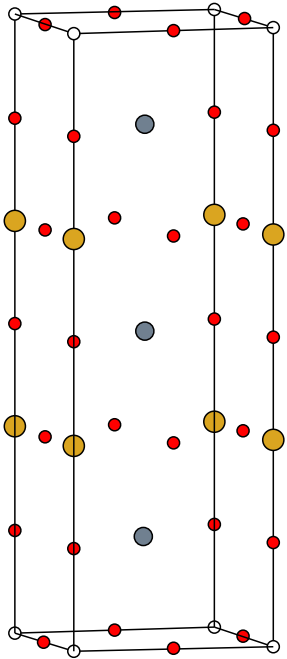}
\end{center}
\vspace{10pt}
\caption{The [001]-ordered Pb(Zr$_{1/3}$Ti$_{2/3}$)O$_{3}$ supercell.
Tables I and II give the
corresponding atomic positions
in the paraelectric and ferroelectric phases.}
\label{fig1}
\end{figure}

We focus on the ordered structure shown in
Fig.~1 and exhibiting a Ti composition equal to 2/3.
The B-site ordering of this supercell consists of one Zr plane
alternating with two Ti planes along the [001] direction.
We perform local-density approximation
(LDA) calculations on
this supercell using the Vanderbilt ultrasoft-pseudopotential scheme 
\cite{David}, and including the semicore shells for {\it all} the metals 
considered.
Specifically, the Pb $5d$, $6s$ and $6p$, the Zr $4s$, $4p$, $4d$ and $5s$,
the Ti $3s$, $3p$, $3d$ and $4s$, and the O $2s$ and $2p$ electrons are treated
as valence electrons.
We choose the plane-wave cutoff to be 25 Ry and use the 
Ceperley-Alder exchange and correlation \cite{Ceperley} as parameterized
by Perdew and Zunger \cite{Perdew}. The first-principles calculations 
throughout this work are performed using a (6,6,2) Monkhorst-Pack 
mesh \cite{Monkhorst}.
Further technical details of the procedure used in the present
study can be found in Ref.~\cite{Domenic}.    
   
In fact, we perform two different calculations 
corresponding to two different symmetries of the structure shown in
Fig.~1: (1) using a centrosymmetric cell 
(i.e., exhibiting an inversion symmetry about the central Pb atom); and 
(2) using a ferroelectric cell (i.e., relaxing the inversion symmetry
constraint).
Results of calculation (1) identify the alloying effects on various
physical properties, while 
comparison of (1) with (2) allows us to isolate the ferroelectric effects
on those properties.

The lattice parameter, the axial ratio c/a and the atomic positions along
the [001] (compositional) direction are optimized in calculation (1) 
by minimizing the total energy and the Hellmann-Feynman forces, the latter
being converged to within 0.02 eV/\AA.

The ferroelectric experimental ground state of 
Pb(Zr$_{1/3}$Ti$_{2/3}$)O$_{3}$ has the
tetragonal P4mm point group and does not present any evidence of (long-range)
B-site ordering.  The material is thus composed of a succession
of equivalent planes of composition Zr$_{1/3}$Ti$_{2/3}$ stacked
along the tetragonal direction.
To mimic this situation, we have chosen the ferroelectric direction
in the supercell of Fig.~1 to lie along the [100] direction, rather
than along the compositionally-modulated [001] direction.
In our ferroelectric supercell, two axial ratios thus exist.  These
are the ``ferroelectric-related'' $a_1/a$ and the ``ordered-related''
$c/a$, where $a_1$, $a$, and $c$ are the lengths of the supercell lattice
vectors along the [100], [010], and [001] directions, respectively.
We are thus dealing with a 
P2mm orthorhombic ferroelectric supercell rather than with a 
P4mm tetragonal ferroelectric supercell.
However, in order to be as close as possible to the experimental 
situation, we will keep the $c/a$ ratio as equal to the ideal value of 3.
In this case, we shall refer to our ferroelectric cell as
``quasi-tetragonal'' along [100] (i.e., tetragonal as regards the
axial ratios, although true tetragonal symmetry is broken down to
orthorhombic by the B-plane ordering in the [001] direction).
The lattice parameter $a$, the axial ratio $a_1/a$, and the atomic positions 
along the [100] and [001] directions are then optimized in calculation (2) 
by minimizing the total energy and the Hellmann-Feynman forces (again
to within a tolerance of 0.02 eV/\AA\ for the forces).

The determination of the electronic ground state in 
calculations (1) and (2) is used to investigate the
ferroelectric effects on the bond length distribution
and on the chemical bonding in PZT with high Ti content.
The effective charges in each case (i.e., 
in both non-centrosymmetric and centrosymmetric cells)
will then be calculated from the polarization differences between 
the ground state and slightly
distorted structures, following the procedure introduced in
Ref.~\cite{Domenic2} and intensively used in Ref.~\cite{Zhong}.       

\section*{Results}

\subsection*{Centrosymmetric case}

Optimizing each degree of freedom in the centrosymmetric supercell leads to 
the lattice vectors 
$\vec{a}_{1,c}=a_0[1,0,0]$,
$\vec{a}_{2,c}=a_0[0,1,0]$,
and $\vec{a}_{3,c}=a_0[0,0,2.99]$,
where $a_0$=7.498 a.u. is the lattice parameter.
The renormalized $c/a$ ratio
defined as the actual ratio (i.e., 2.99) divided by the ideal one
(i.e., 3.00) is equal to 0.997 and is thus very close to unity.
For this reason, our centrosymmetric supercell
can be referred as ``quasi-cubic'', which is consistent with
the fact that the experimental paraelectric phase of PZT is cubic.

\begin{table}
\caption{Structural  relaxations and effective charges for the [001] 
centrosymmetric supercell. The $\Delta z$ are the [001] atomic
displacements of the non-polar structure with respect to the ideal
ordered structure associated with
$\vec{a}_{1,c}$, $\vec{a}_{2,c}$ and $\vec{a}_{3,c}$.
$Z_{xx}$ and $Z_{zz}$ are the effective charges along the 
[100] and [001] direction, respectively. }
\label{Table I}
\begin{tabular}{lccccccc}
\tableline
\tableline
& \multicolumn{3}{c}{Relaxed positions} & Displacements &
  \multicolumn{2}{c}{Effective charges} \\
Atoms   &~~$x$ (a.u.)&~~$y$ (a.u.)&~~$z$ (a.u.)&~~$\Delta z$ (a.u.)  & $Z_{xx}$ & $Z_{zz}$\\
\tableline
 Pb1 & 3.749 &3.749 &-7.194   &~0.279   &~3.90 &~4.04  \\
 Pb2 & 3.749 &3.749 &~0.000   &~0.000   &~3.88 &~3.53\\
 Pb3 & 3.749 &3.749 &~7.194   &-0.279   &~3.90 &~4.04\\
 Ti1 & 0.000 &0.000 &-3.643   &~0.093   &~6.77 &~6.65 \\
 Ti2 & 0.000 &0.000 &~3.643   &-0.093   &~6.77 &~6.65\\
 Zr1 & 0.000 &0.000 &11.210   &~0.000   &~6.33 &~6.69 \\
 O1  & 0.000 &0.000 &-7.222   &~0.251   &-2.58 &-5.39\\
 O2  & 3.749 &0.000 &-3.638   &~0.099   &-5.58 &-2.34 \\
 O3  & 0.000 &3.749 &-3.638   &~0.099   &-2.72 &-2.34\\
 O4  & 0.000 &0.000 &~0.000   &~0.000   &-2.53 & -5.57 \\
 O5  & 3.749 &0.000 &~3.638   &-0.099   &-5.58 &-2.34\\
 O6  & 0.000 &3.749 &~3.638   &-0.099   &-2.72 &-2.34\\
 O7  & 0.000 &0.000 &~7.222   &-0.251   &-2.58 &-5.39\\
 O8  & 3.749 &0.000 &11.210   &~0.000   &-5.17 &-2.94\\
 O9  & 0.000 &3.749 &11.210   &~0.000   &-2.33 &-2.94\\
\tableline
\end{tabular}
\end{table}

The relaxed atomic positions and the effective charges in this non-polar 
structure are given in Table I.
It can be seen from Table I that (i) the Pb and
O atoms lying between the Zr and Ti planes (i.e., Pb1, O1, Pb3, and O7)
move significantly towards the Ti planes; and (ii) the Ti and
O atoms belonging to the Ti planes (i.e., Ti1, O2, O3, Ti2, O5, and O6)
move very slightly towards the central mirror (PbO) plane.
These atomic motions lead to shortened Ti--O bonds and lengthened
Zr--O bonds.  For example, the Ti1--O1 bond length shrinks to 1.89 \AA,
while Zr1--O7 enlarges to 2.11 \AA, to be compared with
the unrelaxed B--O bond length of 1.98 \AA\ in the ideal structure.
As a matter of fact, the appearance of several different bond lengths
associated with the mixed
sublattice seems to be a general feature of alloying, and has also
been observed and predicted in zinc-blende, wurtzite and rocksalt alloys
\cite{Mikkelsen,Frenkel,Martins,Nicola,Anderson,Laurent1}.

Alloying effects in the present ordered [001] structure also lead
to a change in the lengths of the Pb--O bonds. For example, the
Pb3--O bonds can be decomposed into three different groups:
shorter Pb3--O bonds (e.g., Pb3--O5 equal to 2.73 \AA),
roughly unrelaxed Pb3--O bonds (e.g., Pb3--O7 equal to 2.80 \AA),
and long Pb3--O bonds (e.g., Pb3-O8 equal to 2.91 \AA).
The three groups are populated in the ratio 4:4:4.  
Thus, alloying has some significant effects on the B--O bonds 
($\sim$4.5\% change in bond lengths), and to a smaller extent,
on the Pb--O bonds ($\sim$2.5\% change).

\begin{table}
\caption{Structural relaxations and effective charges for the  
non-centrosymmetric supercell.}
\label{Table II}
\begin{tabular}{lccccccccc}
\tableline
\tableline
& \multicolumn{3}{c}{Relaxed positions (a.u.)}
& \multicolumn{3}{c}{Displacements (a.u.)}
& \multicolumn{2}{c}{Effective charges} \\
Atoms &$x$ &$y$ &$z$ &$\Delta x$ &$\Delta y$ &$\Delta z$ &$Z_{xx}$ &$Z_{zz}$\\
\tableline
 Pb1 & 3.547 & 3.731 & -7.250 & -0.334 &0.000 &~0.211 &~3.17 &~3.92  \\
 Pb2 & 3.551 & 3.731 & ~0.000 &  -0.331 & 0.000 &~0.000 &~3.37 &~3.56  \\
 Pb3 & 3.547 & 3.731 & ~7.250 & -0.334 &0.000 &-0.211 &~3.17 &~3.92  \\
 Ti1 & 0.015 & 0.000 & -3.630 &~0.015 & 0.000 &~0.101  &~5.38 &~5.81 \\
 Ti2 & 0.019 & 0.000 & ~3.630 &~0.015 & 0.000 &-0.101 &~5.38 &~5.81  \\
 Zr1 & 0.118 & 0.000 & 11.192 &~0.118 & 0.000 &~0.000 &~6.06 &~6.06 \\
 O1 & 0.628 & 0.000 & -7.223 &~0.628 & 0.000 &~0.238 &-2.15 &-4.80  \\
 O2 & 4.432 & 0.000 & -3.637 &~0.551 & 0.000 &~0.094 &-4.56 &-1.95  \\
 O3 & 0.605 & 3.731 & -3.634 &~0.605 & 0.000 &~0.097 &-2.16 &-2.59  \\
 O4 & 0.656 & 0.000 & ~0.000 &~0.656 & 0.000 &~0.000 &-2.10 &-4.92  \\
 O5 & 4.432 & 0.000 & ~3.637   &~0.551 & 0.000 &-0.094 &-4.56 &-1.95  \\
 O6 & 0.605 & 3.731 & ~3.634   &~0.605 & 0.000 &~0.097  &-2.16 &-2.59  \\
 O7 & 0.628 & 0.000 & ~7.223   &~0.628 & 0.000 &-0.238 &-2.15 &-4.80  \\
 O8 & 4.201 & 0.000 & 11.192   &~0.320 & 0.000 &~0.000  &-4.62 &-2.63  \\
 O9 & 0.779 & 3.731 & 11.192   &~0.779 & 0.000 &~0.000  &-2.07 &-2.90  \\
\tableline
\end{tabular}
\end{table}

The Born effective charges for our PZT supercell are detailed in Table I.
They exhibit the same trends as in cubic bulk PbTiO$_{3}$ and
PbZrO$_{3}$ compounds \cite{Zhong}: large values of about +4.0 for Pb atoms;
large values around +6.5 for the B atoms; and two sets 
of values for the oxygen atoms, either close to -5.5 for oxygen atoms moving 
parallel to the B--O--B chain,
or close to -2.5 for oxygen atoms moving perpendicular to these chains.
The large values of the effective charges for B and O
atoms are due to a (weak) hybridization between the B $d$ and O $2p$ orbitals
\cite{Zhong,Posternak}.
It is interesting to note that along the [001] axis, the effective charge
of Ti is very similar to that of Zr, while 
the difference between these two effective charges in the bulk parent
compounds is larger than 1.0 (i.e., 7.06 {\it vs.} 5.85 for Ti and Zr
respectively, according to Ref.~\cite{Zhong}).
One can also point out that the effective charge along [001]
for atom O7 sitting
between the Ti and Zr atoms is -5.39, i.e., very close
to the average value -5.32 of the corresponding oxygen effective charges
in the bulk parents (-5.83 and -4.81 for PbTiO$_3$ and PbZrO$_3$
respectively, according to Ref.~\cite{Zhong}).

\subsection*{Ferroelectric effects}

We now turn to a consideration of the {\it ferroelectric} effects on
bond-length distributions and effective charges.
Optimizing each degree of freedom in the
non-centrosymmetric cell previously described yields the following
lattice vectors in atomic units: 
$\vec{a}_{1,nc}=a_0'[1.040,0,0]$,
$\vec{a}_{2,nc}=a_0'[0,1,0]$,
and $\vec{a}_{3,nc}=a_0'[0,0,3.00]$,
where the lattice parameter is $a_0'$=7.461 a.u., i.e., 0.5\% smaller than
for the centrosymmetric cell.
A similar decrease of the lattice constant of around 0.7\% has also been 
theoretically predicted when going from the paraelectric cubic phase to the
tetragonal ferroelectric phase of the bulk PbTiO$_{3}$ compound 
\cite{Domenic,Alberto}.
By looking at  $\vec{a}_{1,nc}$, we also notice that 
our calculation predicts 
a ``ferroelectric-related'' axial ratio $a_1/a$ of 1.040.
This prediction must be very close to the true value in
Pb(Zr$_{1/3}$Ti$_{2/3}$), since recent measurements performed 
on Pb(Zr$_{1-x}$Ti$_{x}$) films \cite{Zhu} for $x$=0.6
found a value of 1.035 for this ratio, to be compared
with values of 1.064 and 1.02 for $x=1$ and $x\simeq0.5$
respectively \cite{Lines}.
The optimized non-centrosymmetric cell has an energy
that is 0.15 eV/5-atom-cell lower than that of the
optimized centrosymmetric cell. This is consistent
with the fact that the experimental ground state of 
Pb(Zr$_{1/3}$Ti$_{2/3}$) is ferroelectric and tetragonal rather than
paraelectric and cubic.
The relaxed atomic positions and the effective
charges in the non-centrosymmetric structure are shown in Table II.
The quantities $\Delta x$, $\Delta y$, and $\Delta z$ are the 
[100], [010] and [001] atomic
displacements of the ferroelectric structure with respect to the ideal
ordered structure associated with
$\vec{a}_{1,nc}$, $\vec{a}_{2,nc}$ and $\vec{a}_{3,nc}$.
$Z_{xx}$ and
$Z_{zz}$ are the effective charges along the [100] and [001] directions,
respectively. 

A comparison of Tables I and II leads to the following observations.
(i) The atomic displacements along the [001] direction are quite comparable
between the centrosymmetric and the non-centrosymmetric phases. (ii)
The ferroelectricity in tetragonal PZT is mainly characterized  by
the very large displacement of oxygen atoms along the tetragonal
[100] direction (as in tetragonal PbTiO$_{3}$ bulk), the large
displacement of Pb atoms along the [$\bar{1}$00] direction, and by the
slight displacement of Zr atoms along the [100] direction.

This ferroelectric atomic relaxation yields two different Ti--O bond
lengths along the [100] direction: 
a very long bond of length 2.33 \AA, which is even longer than the 
longest Zr--O bond (2.16 \AA); and a very short bond of length 1.55 \AA.
This very short Ti--O bond is much shorter than the shortest Zr--O bond
of 1.95 \AA, and is even much shorter than the shortest Ti--O bond
of 1.78 \AA\ occurring in tetragonal ferroelectric PbTiO$_{3}$.

Ferroelectricity also leads to  a drastic change in the Pb--O bonds.
There are now some very short Pb--O bonds with an average length of 2.51 \AA,
``normal'' Pb--O bonds with an average length of 2.84 \AA, 
and very long Pb--O bonds with an average length of 3.26 \AA.
As in the centrosymmetric cell, the population ratio between these
three groups is again 4:4:4.
However, the oxygens exhibiting the shortest Pb--O bonds
now share a common (100) plane, while they share a common (001) plane
in the non-polar structure. 
Pair-distribution function analysis of recent pulsed neutron powder diffraction
measurements on ferroelectric PZT alloys clearly confirms the existence
of these three different groups \cite{Egami}. The experimental average
value of the three different Pb--O bond lengths is $\sim$2.5 \AA,
$\sim$2.9 \AA, and
$\sim$3.4 \AA, i.e., in excellent agreement with our predictions.

Comparing Table I and Table II also indicates that ferroelectricity
leads to a striking decrease of the Born effective charges.
The most spectacular decrease occurs for the atoms exhibiting a large change
in their bond lengths. As a matter of fact, the effective
charges along the [100] direction for atoms Ti1, Pb3 and O6
all decrease by
$\sim$20\% with respect to the centrosymmetric case. In consequence,
the effective charges of the Ti atoms and Pb atoms are reduced by 1.4 and 0.7, 
respectively, relative to their non-polar values.
Previous theory has shown that
a change of the effective charge by more than one
unit of `e' is indeed not unusual in going from the cubic to 
tetragonal ferroelectric
phase in perovskite compounds \cite{Ghosez,Wang}. 
The most extreme example appears to be for Nb in
KNbO$_3$, where the effective charges change from 9.67 to 7.05
in the direction parallel to the tetragonal axis.

To further understand the ferroelectric effects in PZT, Fig.~2 
compares the electronic charge density in the (B,O) planes for the
\begin{figure} [t!] % fig 2
\begin{center} \leavevmode
\epsfysize 3.6truein
\epsfbox{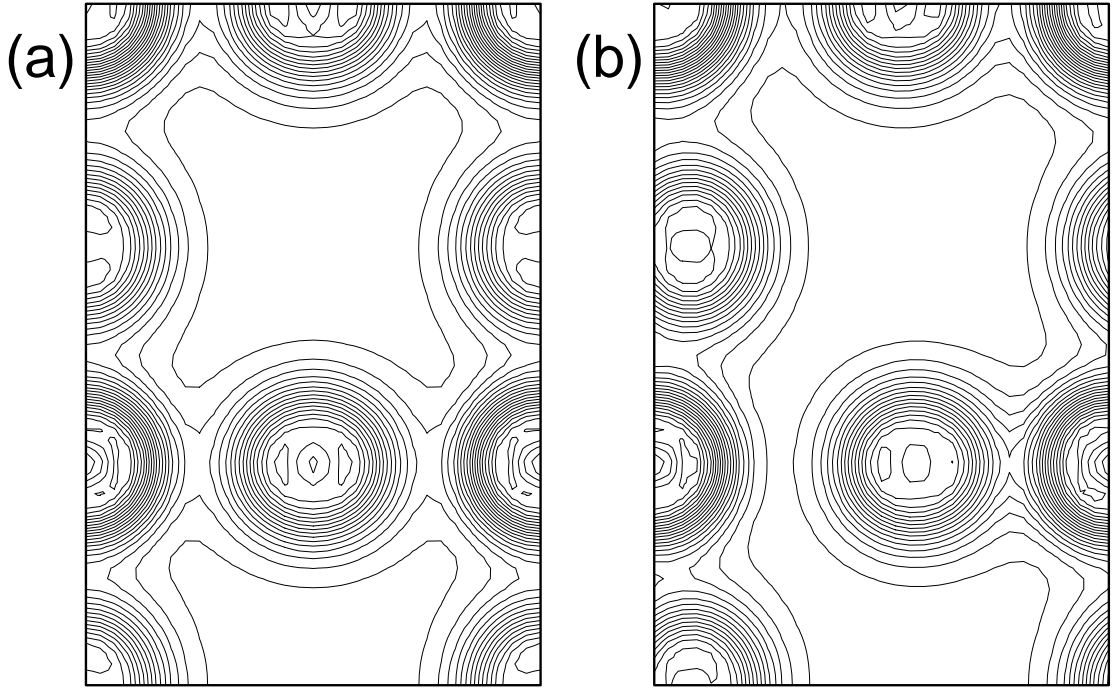}
\end{center}
\vspace{10pt}
\caption{Electronic charge density plotted in the (B,O) plane for
(a) paraelectric, and (b) ferroelectric,
Pb(Zr$_{1/3}$Ti$_{2/3}$)O$_{3}$ supercells.
Only the upper half of the supercell is shown; the horizontal
and vertical axes lie along [100] and [001] respectively.
The sequence of atoms appearing at the left and right edges
is Zr (top), O, Ti, O (bottom); and the remaining middle
atoms are oxygens.}
\label{fig2}
\end{figure}
centrosymmetric and non-centrosymmetric cases.
Figure 3 shows a similar comparison but in the (Pb,O) planes.
Figure 2 indicates that ferroelectricity in PZT leads (i) to a chemical 
breaking of some Ti--O bonds which generates the long Ti--O bonds of 
2.33 \AA, and (ii) to the formation of strong covalency between Ti and
O, which is the cause of the very short Ti--O bonds of 1.55 \AA.
In fact, we also found similar behavior for the electronic charge density
in the (Ti,O) plane for cubic and tetragonal lead titanate (PT).
Thus, as in bulk PT \cite{RCohenN}, the 
formation of ferroelectricity in tetragonal PZT leads 
to an enhancement of hybridization between Ti $3d$ and O $2p$ orbitals.  
Interestingly, we don't observe in Fig.~2 any breaking of Zr--O bonds 
nor the formation of strong covalent Zr--O bonds. The different chemical 
behavior between Zr and Ti may perhaps be the cause of the difference in
ground states exhibited by the corresponding bulk parents
(antiferroelectric and orthorhombic for PbZrO$_{3}$ {\it vs.}
ferroelectric and tetragonal for PbTiO$_{3}$).
The striking feature of Fig.~3 is the formation of covalent chains between
Pb and O atoms, which is the cause of the very short Pb-O bonds of  
2.5 \AA.
We also found similar trends in the (Pb,O) planes of paraelectric
and ferroelectric PT.
Thus, as in bulk PT \cite{RCohenN},
the hybridization between Pb $6s$ and O $2p$ orbitals plays an 
important role in the ferroelectric behavior of tetragonal PZT.

\begin{figure} [!t] % fig 3
\begin{center} \leavevmode
\epsfysize 3.6truein
\epsfbox{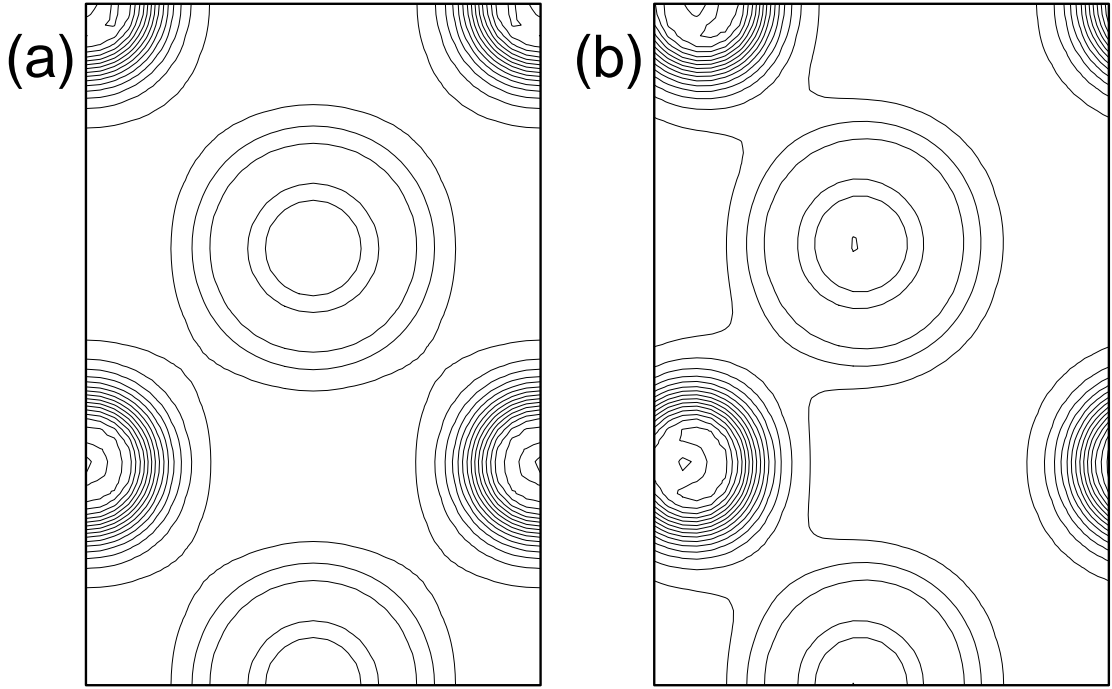}
\end{center}
\vspace{10pt}
\caption{Electronic charge density plotted in the (Pb,O) plane for
(a) paraelectric, and (b) ferroelectric,
Pb(Zr$_{1/3}$Ti$_{2/3}$)O$_{3}$ supercells.
Only the upper half of the supercell is shown; the horizontal
and vertical axes lie along [100] and [001] respectively.
Atoms appearing at the left and right edges are O;
atoms in the middle are Pb.}
\label{fig3}
\end{figure}
\section*{Conclusions}
Using 15-atom supercells and Vanderbilt ultrasoft pseudopotentials 
within the local-density approximation, we investigated alloying and 
ferroelectric effects on the bond lengths, chemical bonding and effective 
charges in lead zirconate titanate alloys (PZT).

Our principal findings are as follows.

(i) The centrosymmetric PZT
alloy is mainly characterized by two sets of B--O bonds (shorter
Ti--O bonds {\it vs.}\ longer Zr--O bonds), while the Pb--O bonds
differ only slightly (by $\sim$2.5\%) from the ideal structure.

(ii) Allowing ferroelectricity in PZT alloys has two striking chemical effects:
enhancement of hybridization between Ti $3d$ and O $2p$ orbitals, and
hybridization between Pb $6s$ and O $2p$ orbitals. 

(iii) These chemical and ferroelectric effects
lead to the formation of very short covalent Ti--O bonds
while breaking other Ti--O bonds, and give rise to the formation of
covalent chains of very short Pb--O bonds. 

(iv) The atoms engaged in covalent
bonding exhibit a striking decrease of their effective charges by $\sim$20\%
relative to the paraelectric phase.   

\section*{Acknowledgments}
This work is supported by the Office of Naval Research grant N00014-97-1-0048.
We thank Professor T. Egami for helpful discussions and for communicating
his results with us.

\end{document}